\documentclass[twocolumn,prl,aps]{revtex4}

\usepackage[dvips]{graphicx}
\usepackage[dvips]{graphics}

\begin{document}

 \title{Observation of Magnetic Order in a ${\rm YBa_2Cu_3O_{6.6}}$ Superconductor}

 \author{H. A. Mook$^{1,*}$,  Y.~Sidis$^2$, B. Fauqu\'e$^2$, V. Bal\'edent$^2$, and P.~Bourges$^{2}$}

\affiliation{
$^1$ Neutron Scattering Science Directorate, Oak Ridge National Laboratory, Oak Ridge, Tennessee 37831-6393, USA\\
$^2$ Laboratoire L\'{e}on Brillouin, CEA-CNRS, CEA-Saclay, 91191 Gif sur Yvette, France
}

\begin{abstract}
Polarized beam neutron scattering measurements on a highly perfect crystal of  ${\rm YBa_2Cu_3O_{6.6}}$ show a distinct magnetic transition with an onset at about 235K, the temperature expected for the pseudogap transition. The moment is found to be about 0.1 $\mu_B$ for each sublattice and have a correlation length of at least 75 \AA. We found the critical exponent for the magnetic neutron intensity to be 2$\beta$ =0.37$\pm$ 0.12. This is the proper range for the class of transition that has no specific heat divergence possibly explaining why none is found at the pseudogap transition. 
\end{abstract}

\maketitle

Superconductivity produces a gap in the quasiparticle spectrum which for conventional materials disappears as the temperature is increased to $T_c$, where the superconducting electron pairs are no longer bound together. However, for the underdoped cuprate materials a pseudogap measured by a number of techniques appears at a temperature T*$>T_c$, with T* increasing as $T_c$ gets smaller. This pseudogap associated with T* is one of the most puzzling and important attributes of the cuprate superconductors\cite{workshop,timusk} Indeed since superconductivity originates from the pseudogap state it is this state that has to be understood to determine the mechanism for high temperature superconductivity. One possible origin of the pseudogap is to postulate that phase incoherent pairs are established as the material is cooled through the pseudogap temperature T*, with superconductivity developing at a lower temperature $T_c$ when phase coherence is established\cite{emery} This picture has been studied by the Nernst effect\cite{xu}, which shows the presence of vortex-like excitations in the pseudogap phase. However, upon warming the Nernst\cite{xu} effect disappears well before T* is reached suggesting that well defined preformed pairs may not be present at temperatures as high as T*.

Completely different approaches\cite{cmv1,cmv2,ddw} to the problem postulated a state with broken symmetry that displayed a pattern of circulating currents (CC phase) in the a-b plane. As the sample is cooled this state appears at T* and for near optimal doping ends at a quantum critical point. The present study considers a magnetic signal found at certain (h,k,l) positions in the reciprocal lattice. These would correspond to the phases considered by Varma\cite{cmv1,cmv2} which preserve the translational symmetry of the lattice. Since translational symmetry is not broken the signal to observe the phases occur at the Bragg positions of the unit cell which means they sit on top of the much more intense nuclear scattering. A very sensitive polarized beam experiment is thus needed to observe them. Fauqu\'e, et al\cite{fauque} have undertaken polarized neutron scattering measurements to search for this phase, at the expected positions for several underdoped ${\rm YBa_2Cu_3O_{6+x}}$samples. They observed an increase in polarized neutron scattering upon cooling, which represents a magnetic signal, which could be associated with the pseudogap temperature T*.  We have undertaken similar polarized neutron measurements on a well characterized single crystal of underdoped ${\rm YBa_2Cu_3O_{6.6}}$ ($T_c$=63 K). Since the crystal has highly perfect Ortho II order, a very sharp superconducting transition shown in Fig.1b, and is quite large (25 grams) we hoped a better defined magnetic transition could be observed and that additional important information about the magnetic state could be obtained.  

\begin{figure}
 \includegraphics[height=8cm,width=8cm]{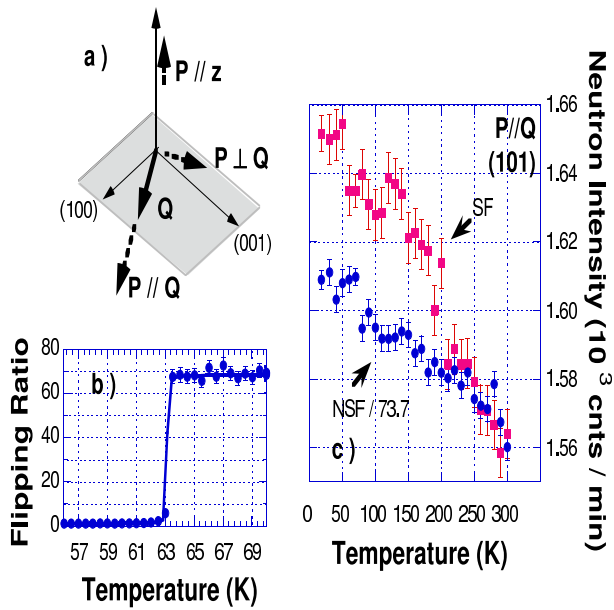}
\begin{center}
\caption{(color online) The scattering diagram used in the experiment, the superconducting transition, and raw data acquired for the (1, 0, 1) reflection.  Panel a shows the scattering diagram where Q is positioned for the (1, 0, 1) reflection. P//Q and P$\perp$Q are in the scattering plane and are two of the polarization states used. The third state is (P//z) where z is perpendicular to the scattering plane. b shows the superconducting transition obtained by measuring the polarization transmitted through the crystal in zero applied field. The transition is very sharp for an underdoped crystal of the size used. c shows a polarized measurement made in the P//Q arrangement where the red squares show the SF magnetic scattering and the blue points are the NSF non magnetic scattering normalized to the SF scattering at room temperature. The distinct jump in the SF data below 230K shows the magnetic transition.  } 
\label{fig1}
\end{center}
\end{figure}

We aligned our YBCO (x=6.6) sample in the [H,0,L] scattering plane, which is the orientation used by Fauqu\'e, et al\cite{fauque} in their experiments (Fig. 1a). Since the sample is twinned we do not differentiate between the [0,K,L] and [H,0,L] directions. All the polarized neutron diffraction measurements were collected on the 4F1 triple-axis spectrometer at the Laboratoire L\'eon Brillouin, Saclay, France. Our polarized neutron diffraction setup is similar to that originally described in ref. \cite{fauque} with a polarized incident neutron beam at Ei =13.7 meV obtained with a polarizing super-mirror (bender) and a Heusler analyzer. A pyrolytic graphite filter was used in the incident beam before the bender to reduce background neutrons. Here the scattering wave vector Q = ($Q_x, Q_y, Q_z$) in \AA$^{-1}$ has been labeled as (H, K, L) = ($Q_x 2\pi/a, Q_y 2\pi/b, Q_z 2\pi/c$) in reciprocal lattice units. The analyzer size was reduced to improve the polarization efficiency. The standard polarized technique was employed where a neutron spin-flipper was placed before the sample to reverse the polarization of neutrons. The flipping ratio is defined as the ratio of the non-spin flip (NSF) neutron intensity, where the polarization is kept the same, over the spin flip (SF) neutron intensity where the spin of neutrons is flipped. All magnetic scattering stems from a moment that lies in the plane $\perp$ to Q. A magnetic guide field ($\sim$10 Oe) at the sample position was controlled by coils to establish the polarization in the desired direction. We made measurements for three different polarization configurations.  Two of these are with the polarization P in the scattering plane, P$\perp$Q and P//Q. The third is with P $\perp$ to the scattering plane which we will denote by P//z as shown in Fig. 1a. For P//Q, all the magnetic scattering will be spin-flip (SF) and the magnetic moment producing the scattering is directed anywhere in the plane $\perp$ Q. This last condition also applies for the other two polarization directions plus the extra condition that the moment lies in the plane  $\perp$ to P for each case. The polarization is very high for a neutron experiment, with a flipping ratio R=NSF/SF being about 75. 

	Fig.1c shows raw data for the (1, 0, 1) reflection using the P//Q polarization configuration.  Here the non-spin-flip scattering which is non-magnetic has been matched at 300K to the spin-flip SF scattering which contains possible magnetic scattering. The NSF scattering shows how the Debye-Waller factor affects the data or how the SF scattering would look if there were no magnetic transition. The SF scattering follows the NSF scattering until about 230K at which point it rises rapidly above the NSF scattering demonstrating the transition to a magnetic state.  That observation confirms the previous measurement of Fauqu\'e et al \cite{fauque}  of a magnetic order associated with the pseudogap state. However, the transition is sharp relative to the previous report which showed a near linear increase with temperature. 

		A crucial part of the measurement is determining the temperature dependence of the NSF/SF background. We ignored this in Fig. 1b but to get the correct size of the magnetic signal this must be known. The magnetic signal drops off as the momentum transfer is increased so that the background may be obtained by making a measurement at a high Q reflection where the magnetic form factor considerably reduces the signal.  We chose the (2, 0, 1) reflection, which is not so far out that the spectrometer configuration is not greatly changed, but far enough out that magnetic scattering is expected to be small.  

\begin{figure}[t]
 \includegraphics[height=8cm,width=8cm]{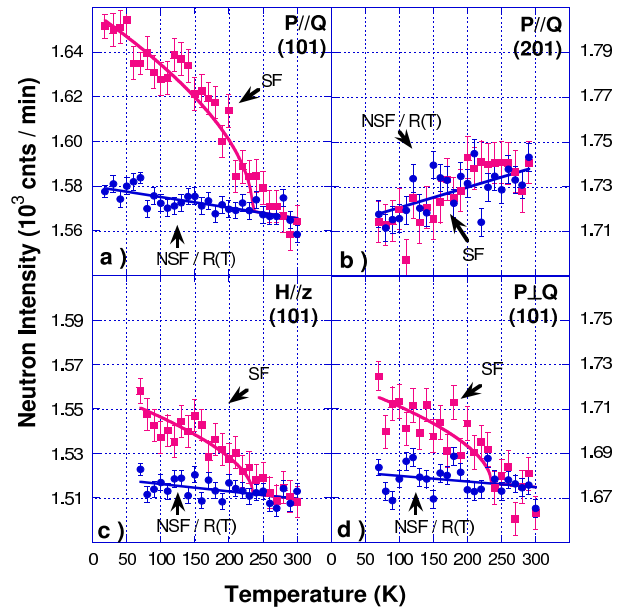}
\begin{center}
\caption{Measurements for the three polarization configurations on the (1, 0, 1) reflection and for the P//Q configuration on the (2, 0, 1) reflection. The fit for the (1, 0, 1) reflection shown in panels a, c, d by the red lines is given by $T<T_m, I=BG+((T_m-T)/T_m)^2$, $T>T_m$  I=BG,  BG=A+BT. The best fit is for Tm=235$\pm$15K and 2$\beta$ =0.37$\pm$0.12 where BG is a linear background given by the blue line. BG is obtained by dividing the NSF intensity by the temperature dependence of the flipping ratio, R(T). R(T) is found by a fit in panel b of the ratio NSF/SF of the Bragg peak (2, 0, 1) as R(T)=R(300 K){1+0.02(1-T/300)}.
} \label{fig2}
\end{center}
\end{figure}

	Fig.2 shows the results of measurements in the three polarization configurations where the background is included in the data analysis. The background obtained from the (2, 0, 1) reflection is shown in Fig. 2b. This background is in good agreement with that obtained by the sum rule on the polarization states in which data from the P//Q configuration must be equal to the sum of that taken in the P$\perp$Q and P//z conditions. Such a good agreement, also reported in ref. \cite{fauque}, shows that the additional scattering present below around 230 K at (1, 0, 1) in the SF channel is absent (or non-observable within errors) at the (2, 0, 1) reflection. Figs. 2a, c, and d show the results of the measurements in the three polarization conditions where the background, denoted in blue, is shown for all the measurements. From these results the moment direction may be obtained and is found to be at a position of 55$\pm$7 degrees from the c axis. 

\begin{figure}[t]
 \includegraphics[height=8cm,width=8cm]{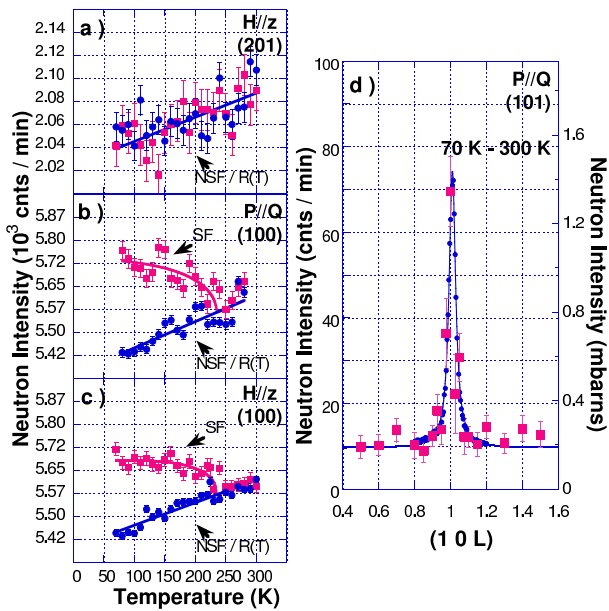}
\begin{center}
\caption{Measurements for two polarization configurations on the (1, 0, 0) reflection and for the P//z configuration on the (2, 0, 1) reflection and a scan through (1, 0, 1) along c*. In panel a the background given by the NSF scattering corrected by the temperature dependence of the flipping ratio R(T) (blue circles) is compared to the SF scattering (red squares) for the Bragg (2, 0, 1). Panels b and c show the SF scattering (red squares) and NSF/R(T) (blue points) for the P//Q and P//z  polarization directions for (1, 0, 0) (R(T)=R(300 K){1+0.06(1-T/300)} has been determined from Fig.  3b). The difference between both curves shows the magnetic scattering. d shows a scan along L for the (1, 0, 1) position taken in the P//Q polarization state of the magnetic scattering (red squares), obtained by the temperature difference of the SF scattering where NSF/R(T) has been subtracted at each temperature (R(T)=R(300 K){1+0.02(1-T/300)} as in Fig. 2). The blue points show a NSF non-magnetic scan of the Bragg peak normalized to the magnetic intensity.  Results of panels a, b and c have been obtained with the same experimental setup whereas the L-scan of panel d has been measured with the same experimental conditions as Fig 2. } \label{fig3}
\end{center}
\end{figure}

	Fig. 3 gives results obtained from the (1, 0, 0) reflection. This reflection has a considerably larger nuclear cross section than the (1, 0, 1) so that it is more difficult to obtain accurate data. Because the counting errors are larger it is difficult to determine the magnetic transition as accurately as for the (1, 0, 1), but the two results are consistent with each other. Fig. 3 a and b show data taken with different polarization conditions at (1, 0, 0) that make it possible to calculate the moment direction, which is found to be 35$\pm$7 degrees from the c axis. The errors of $\pm$7 obtained at both Bragg reflections are derived from statistical errors only and improve the earlier result of 45$\pm$20 degrees\cite{fauque}. On the other hand such a difference is predicted by a model\cite{vivek} that considers moments from both CC and spin moments, the latter being induced through the spin-orbit coupling. The model gives a smaller angle for the (1, 0, 0) reflection as it is found. Fig. 3d shows a scan along the L direction at the (1, 0, 1) position. The magnetic scattering is obtained by taking a temperature difference of the SF scattering for the P//Q polarization condition. The blue points represent the non-magnetic NSF Bragg scattering (i.e. the resolution of the spectrometer, here full width at half maximum of 0.013 \AA$^{-1}$) scaled to the same amplitude as the magnetic scattering. Both curves superpose very well meaning that the magnetic peak is limited by the resolution. That suggests that the magnetic state is ordered at long range along c*. From the resolution width, one can determine a lower bound for the magnetic correlation length of 75 \AA. The L-scan shows non-zero background scattering off the peak (Fig. 3d). In contrast to the peak, the polarization analysis shows that this temperature dependent background is non-magnetic. 

	Performing scans in Q space is difficult as the spectrometer must move for each step possibly changing the polarization slightly. This seems to be more of a problem for the H direction rather than for the L direction, possibly because of resolution effects since the crystal mosaic spread is under a degree. We will search for ways to improve this situation, but so far have not had much success. 	.
 
	A major advance here is the observation of magnetic scattering that is not just a gradual increase in intensity but rather a scattering pattern that shows a sharp increase in a narrow temperature range as expected for the transition to a new phase. Furthermore this occurs at the position expected for the pseudogap transition. The requirements to obtain this result are a crystal of exceptional perfection big enough to obtain good counting statistics and an intense polarized beam with a very high polarization. The crystal has been used in other studies that demonstrate its perfection such the discovery of magnetic incommensurate structure in the  ${\rm YBa_2Cu_3O_{6+x}}$ materials\cite{mook} and the sharpness of the resonance peak at ($\pi,\pi$) compared to other crystals\cite{dai}.	

We do not know the origin of the observed magnetism. If we use 0.5 for the square of the (1, 0, 1) magnetic form factor as it has been done in ref. \cite{fauque}, it corresponds to a moment of about 0.1 $\mu_B$ per each sublattice. We found here a magnetic order with a sharp transition near the value of Tm= 235$\pm$15K which respects the translational symmetry of the lattice and whose symmetry corresponds to the one recently proposed by Varma\cite{cmv2}. 

  Of course the possibility that the observed magnetic order is related to the pseudogap is of great interest. Such an interpretation is supported by the work of Xia et al \cite{xia} who also find a time-reversal broken symmetry near the pseudogap temperature from the Kerr effect. We can fit the temperature dependence of the magnetic scattering to obtain the critical exponent would  giving 2$\beta$ =0.37$\pm$ 0.12. Essentially the same number is obtained by plotting the log of the intensity vs log($T_m -T)/T_m$ and obtaining the slope. Since we do not have a high density of points, data down to low temperatures were needed to get the quoted exponent. This should be satisfactory in this case since the transition is likely of the Ising type which in two dimensions has a critical regime of the same order as $T_m$. A  transition with an exponent $\beta$ =0.18 which is in the range between 1/4 and 1/8, has a much reduced specific heat divergence \cite{baxter} possibly explaining why none is observed at the pseudogap transition \cite{cmv3,sudip}. These results provide strong evidence that the pseudogap is associated with a magnetic transition, providing an answer to many of questions about nature of the state from which high temperature superconductivity originates.

Acknowledgements
We appreciate helpful discussions with C. M. Varma,  S.A. Kivelson,  Pengcheng Dai, A. Kapitulnik and S. Chakravarty. Work at Oak Ridge National Laboratory is supported by the US DOE under contract No. DE-AC05-00OR22725 with UT/Battelle.

$^*$To whom correspondence should be addressed; E-mail: mookhajr@ornl.gov

\end{document}